\documentclass[runningheads,citeauthoryear]{apinv2022}
\usepackage{epsfig,cite,graphics}
\usepackage{marvosym} %For astronomical symbols % % %
\usepackage[utf8]{inputenc}

\def\kms{km~s$^{-1}$}  
\newcommand{\msun}{$M_\odot$}

\begin{document}

\title{ $H\alpha$ spectroscopy of the recurrent nova RS Oph during the 
2021 outburst}
\titlerunning{ $H\alpha$ spectroscopy of RS Oph}
\author{R. K. Zamanov\inst{1}, 
K. A. Stoyanov\inst{1}, 
Y. M.  Nikolov\inst{1}, 
T. Bonev\inst{1}, 
D. Marchev\inst{2},
S. Y. Stefanov\inst{1,3}
}
\authorrunning{Zamanov et al. }
\tocauthor{R. K. Zamanov, K. A. Stoyanov, Y. M.  Nikolov, T. Bonev, D. Marchev, S. J. Stefanov} 
% Command tocautor{} is used by the Latex to give author names 
% to the Contents of the volume (automatically generated)
\institute{Institute of Astronomy and National Astronomical Observatory, Bulgarian Academy of Sciences, 
Tsarigradsko Shose 72,  BG-1784, Sofia, Bulgaria
           \and
	   Department of Physics and Astronomy, Shumen University "Episkop Konstantin Preslavski", 
	   115 Universitetska Str., 9700 Shumen, Bulgaria
	   \and 
           Department of Astronomy, Sofia University "St. Kliment Ohridski", 
	   James Bourchier 5, BG-1164 Sofia, Bulgaria	  
        \newline
	\email{rkz@astro.bas.bg  kstoyanov@astro.bas.bg}    }
\papertype{Research report,  Received 24 September 2021 / Accepted .. ...  .......}	
% Papertype can be "Research report", "Review", "Invited lecture", "Conference talk", 
% "Conference poster", "Lecture at scientific seminar", "Summary of dissertation",  etc.
\maketitle

\begin{abstract}
We report spectroscopic observations of the $H\alpha$ emission line
of the recurrent nova RS~Oph obtained between 12 and 23 August 2021 
during the recent outburst.
On the basis of the sharp P~Cyg profile superimposed onto the strong 
$H\alpha$ emission, we estimate that the outflowing velocity of the material surrounding 
RS~Oph is in the range 
32~km~s$^{-1}$ $< V_{out}<$ 68~km s$^{-1}$.
The new GAIA distance indicates that the red giant should be probably
classified in between II and III luminosity class. \\
The spectra are available upon request from the authors and on Zenodo.
 
\end{abstract}

\section*{Introduction}
In the binary systems, the nova outburst is powered by thermonuclear runaway 
on the surface of the white dwarf (see Bode \& Evans 2008 for a comprehensive review). 
For a long period of time, 
the hydrogen-rich material is being accreted from the donor star 
to the  white dwarf and forms an envelope on its surface. 
Once the critical pressure at the base of the accreted layer is reached, 
the thermonuclear runaway is ignited, causing a dramatic brightening 
(e.g. Shafter et al. 2009).
The binary system survives the outburst 
and the mass accumulates once again 
and the nova recurs on a time-scale that depends on the accretion
rate and on the white dwarf mass. 
The objects that experience more than one nova outburst are known as {\it recurrent novae} 
(e.g. Mukai 2015). 
It is believed that all 
classical novae are recurrent with intervals between the outbursts of many centuries. 
% It is confirmed by observations that the nova ejecta enriches the Interstellar Medium with 
% heavy elements and  dust (Gehrz et al. 1998).\\

The 2021 nova outburst of RS~Oph was reported on 2021 August 8.93 by K. Geary (AAVSO Alert Notice 752). The first spectral follow-up observations reveal resemblance of a He/N nova and the 
presence of prominent Balmer lines, and He~I, Fe~II, O~I and N~II 
features with P~Cyg profiles (Taguchi, Ueta \& Isogai 2021; Munari \& Valisa 2021a). Further spectral observations, obtained 2 days after the outburst, indicate an acceleration of the ejecta, reaching 
velocities of about $-4200$~km~s$^{-1}$ and $-4700$~km~s$^{-1}$, 
estimated from the P~Cyg profiles of the H$\alpha$ and H$\beta$ lines 
respectively (Mikolajewska et al. 2021). 
The outburst was detected also  in the radio (Williams et al. 2021), 
X-rays (Shidatsu et al. 2021) and $\gamma$-rays in GeV domain (Cheung, Ciprini \& Johnson 2021)
and TeV domain (Wagner \& H.~E.~S.~S. Collaboration 2021).
% The X-ray observations on 2021 Aug 30  
% suggest a possible beginning of the supersoft source phase (Page 2021).\\

Here we present H$\alpha$ observations of RS~Oph from 2021 Aug 16 to 2021 Aug 23
and discuss (1) the P~Cyg type profile at the top of the emission 
and (2) the absolute magnitude of the red giant.

\section{Observations}
High-resolution optical spectra of RS~Oph are secured with the Coud\`{e} spectrograph attached to the 2m telescope of the 
Rozhen National Astronomical Observatory, located in Rhodope mountains, Bulgaria. The spectra cover 225~\AA\ 
around the H$\alpha$ line with a resolution of 0.11~\AA/pixel. 
The spectra are reduced in the standard way including bias removal, flat-field correction, 
and wavelength calibration using the routines provided in {\sc iraf} (Tody 1993).  
The wavelength calibration is done with Thorium-Argon (Th-Ar) hollow cathode lamp 
and tuned using the telluric lines imprinted in the spectrum (see Fig.~3 in Appendix). 
The FWHM (full width at half maximum) of the Th-Ar lines is 0.4~\AA,  
the FWHM of the telluric lines is 0.5~\AA. The spectra cover the wavelength range from 6450~\AA\  to 6675~\AA.  
The spectra are available upon request from the authors and on Zenodo
(zenodo.org/record/5524465). 

For comparative purposes, we also use 
two spectra obtained with the Echelle spectrograph of the same telescope in 2019 and 2020, 
and one spectrum obtained with an 11 inches Celestron C11 telescope and Lhires III  spectrograph. 
Two observations of the $H\alpha$ emission line are plotted on Fig.~1.
One of them is before the outburst (5 September 2020) 
and the second is in outburst (21 August 2021). 
In outburst, the $H\alpha$ emission is of about 20 times stronger
and 5 times wider.

The journal of observations and the measured parameters  of $H\alpha$ line
are given in Table~1: 
\begin{itemize}
\item  column 1 - date of observation (in format YYYY MM DD HH:MM). The time is the start of the exposure. 
\item  column 2 - telescope and spectrograph; 
\item  column 3 - the exposure time in minutes; 
\item  column 4 - the total equivalent width of $H\alpha$ emission line; 
\item  column 5 - the FWHM  of the $H\alpha$ emission line. 
           This is the FWHM of the strong broad component only. This component is
	    emitted from the expanding envelope.   
\item  column 6 - the wavelength of the diffuse interstellar band DIB 6613, which is used for check of the wavelength calibration; 
\item  column 7 - the heliocentric radial velocity of the absorption part of the P~Cyg profile.
For the spectra 20190718 and 20200905, column 7 is the heliocentric radial velocity
of the central dip, located in between the blue and red peaks. 
\item  column 8 - the heliocentric radial velocity of the emission part of the P~Cyg profile.
\end{itemize} 

%--------------------------------------------------------------------------------
\begin{table}
\caption{Journal of observations and some parameters of $H\alpha$ line. 
In the columns are given as follows: date of observation (in format YYYY MM DD HH:MM), telescope and spectrograph, 
exposure time in minutes,  
the equivalent width of $H\alpha$ emission line, 
FWHM of  the broad component of the $H\alpha$ emission line, 
the heliocentric wavelength of the DIB 6613, radial velocity of the absorption and emission components
of the P~Cyg emission. 
}               
\centering         
\begin{tabular}{c | c | c | c| c| c| c| c }     % 5 columns 
\hline\hline 
  &    &    &    &     &    &    &  \\      
Date-obs  & telescope      & expo       &    EW $H\alpha$    &   FWHM H$\alpha$ & DIB6613  &  RV$_{abs}$  & RV$_{em}$         \\ 
          &                &  [min]     &    [\AA]           &   [\AA]          &   [\AA]  &  [\kms]      & [\kms]            \\
  &    &    &    &     &    &    & \\   
  1       &      2         &   3        &        4           &         5        &    6     &      7       &         8         \\   
\hline     
  &    &    &    &     &    &    & \\       	 
2019 07 18 20:06 &   2m Ech	  &  60 &               & 	           &  6613.369  &  $-70.6 \pm 1.5$  &		      \\
2020 09 05 19:07 &   2m Ech	  &  30 &               &              	   &  6613.318  &  $-62.3 \pm 1.5$  &		      \\
2021 08 12 20:32 &   C11"	  &  30 & $ -780\pm$80  &  $45.8 \pm 0.5$ &	       &		   &		      \\
2021 08 16 20:28 &   2m Coude	  &   3 & $-1350\pm$60  &  $35.7 \pm 0.3$ &  6613.305  &  $-69.7 \pm 2.0$  &  $ -9.1 \pm 2.0$ \\
2021 08 17 20:25 &   2m Coude	  &   5 & $-1300:  $	&  $33.2 \pm 0.4$ &  6613.373  &  $-71.2 \pm 2.0$  &  $-11.6 \pm 2.0$ \\
2021 08 19 20:01 &   2m Coude	  &   5 & $-1650\pm$60  &  $30.0 \pm 0.3$ &  6613.321  &  $-73.2 \pm 2.0$  &  $-12.3 \pm 2.0$ \\
2021 08 20 19:17 &   2m Coude	  &   5 & $-1770\pm$70  &  $28.6 \pm 0.3$ &  6613.391  &  $-72.9 \pm 2.0$  &  $ -7.3 \pm 2.0$ \\
2021 08 21 19:10 &   2m Coude	  &  10 & $-1940\pm$90  &  $27.7 \pm 0.3$ &  6613.342  &  $-74.2 \pm 2.0$  &  $ -9.2 \pm 2.0$ \\
2021 08 22 19:01 &   2m Coude	  &   5 & $-1910\pm$95  &  $26.3 \pm 0.3$ &  6613.353  &  $-74.2 \pm 2.0$  &  $ -9.3 \pm 2.0$ \\
2021 08 23 19:46 &   2m Coude	  &   5 & $-2080\pm$95  &  $24.5 \pm 0.3$ &  6613.296  &  $-77.3 \pm 2.0$  &  $ -4.0 \pm 2.0$ \\
	         & 		  &     &                &                 &            &	 	    &		      \\
\hline                  
\end{tabular}
\label{tab-J}    
\end{table}
%-------------------------------------------------------------
%20190718.hwl.fit	6561.272		-1.55	-70.6	
%20200905.hwl.fit	6561.455		-1.36	-62.3	
%20210816.5d.hwl.fit	6561.292		-1.52	-69.7	
%...	6562.617		-0.2		-9.1
%20210817.02.hwl.fit	6561.259		-1.56	-71.2	
%...	6562.563		-0.25		-11.6
%20210819.02.hwl.fit	6561.215		-1.6	-73.2	
%...	6562.548		-0.27		-12.3
%
%20210820.02.hwl.fit	6561.223		-1.59	-72.9	
%...	6562.657		-0.16		-7.3
%20210821.04.hwl.fit	6561.194		-1.62	-74.2	
%	6562.616		-0.2		-9.2
%20210822.02.hwl.fit	6561.193		-1.62	-74.2	
%	6562.613		-0.2		-9.3
%20210823.02.hwl.fit	6561.127		-1.69	-77.3	
%	6562.729		-0.09		-4.0

\section{Results}

In our data set is visible that 
(1) the FWHM of the the strong $H\alpha$ emission 
originating from the nova ejecta is monotonically decreasing 
from 45.8~\AA\ on 12 August 2021 to 24.5~\AA\ on 23 August 2021 (see Table~1, column 5);
and (2) its EW is increasing.  A more detailed atlas can be found in Munari \& Valisa (2021b). 

\subsection{Absolute V magnitude ($M_V$) of the mass donor}

The light curves of RS Oph during the last 30 years are well documented in AAVSO data. 
In quiescence the V band magnitude of RS~Oph is $10.2 < m_v < 11.3$. 
% During the 2021 outburst it achieved $m_V \ approx 4.5$. 
According to the AAVSO V-band light curve, the maximum brightness of RS~Oph during the outburst is V=4.593~mag. 
The observations suggest an outburst amplitude of 
$\sim$ 6 -- 6.5~mag. 
To calculate the absolute V band magnitude we use the well known formula:
\begin{equation}
M_V = m_V - 3.1 E_{B-V} + 5 \log_{10}(d/10),
\end{equation}
where $m_V$ is the apparent V band magnitude,
$E_{B-V}$ is the interstellar reddening, 
$d$ is the distance in parsecs. 
For RS Oph $E_{B-V}= 0.69 \pm 0.07$ (Zamanov et al. 2018), and $d=2600$ pc ({\it GAIA}~eDR3, 
Gaia~Collaboration et al. 2021).  
Using V=4.593~mag, we obtain an absolute V magnitude at the maximum $M_V \approx -9.69$~mag.

The brightness of the red giant in RS~Oph is $m_v$ $\approx$ 12.26 (Zamanov et al. 2018). 
Using again  Eq.~1, we obtain that
the absolute V-band magnitude of the red giant is $M_V \sim -2.02$~mag.
Following Straizys \& Kuriliene (1981) 
M2III giant has absolute V magnitude  $M_V = -0.9$, while 
for luminosity class II, M2II  --  $M_V =-3.0$.  
It means that the new GAIA distance puts the red giant of RS~Oph in between
luminosity classes III and II.
It is worth noting that with the old value $d=1600$ pc (Bode 1987), 
the red giant would have $M_V \approx -1.0$, in agreement with  M2III spectral type. 

% Using V=4.593~mag, we obtain an absolute V magnitude at the maximum $M_V \approx -9.69$~mag.\\
% The light curves of RS Oph during the last 30 years are well documented in AAVSO data. 
% In quiencecs the V band magnitude of RS~Oph is $10.2 < m_v < 11.3$.
% During the 2021 outburst it achieved $m_V \ approx 4.5$. 
% The observations suggest an outburst amplitude of 
% $\sim$ 6 -- 6.5~mag. 
% Using V=4.593~mag, we obtain an absolute V magnitude at the maximum $M_V \approx -9.69$~mag.\\

%-------------------------------------------------------------------------------------
\begin{figure}
  \vspace{5.6cm}
  \includegraphics{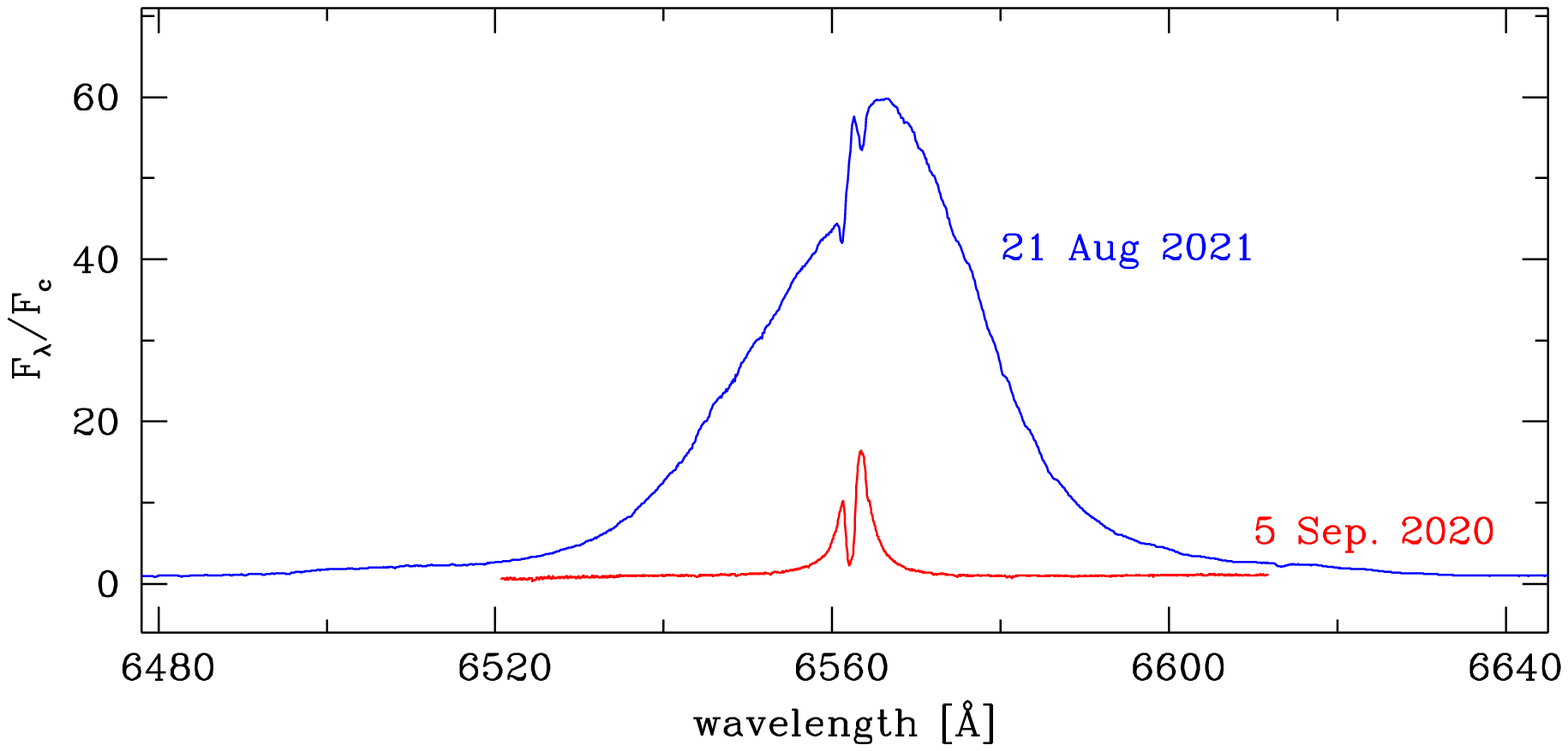}
  \caption[]{A comparison between the H$\alpha$ line profiles at quiescence (2020 Sep 5)   
             and in outburst (2021 Aug 21). 
	     In outburst, the $H\alpha$ emission is $\sim 20$ times stronger
	      and $\sim 5$ times wider. }
   \label{f1}
 \vspace{8.9cm}
 \includegraphics{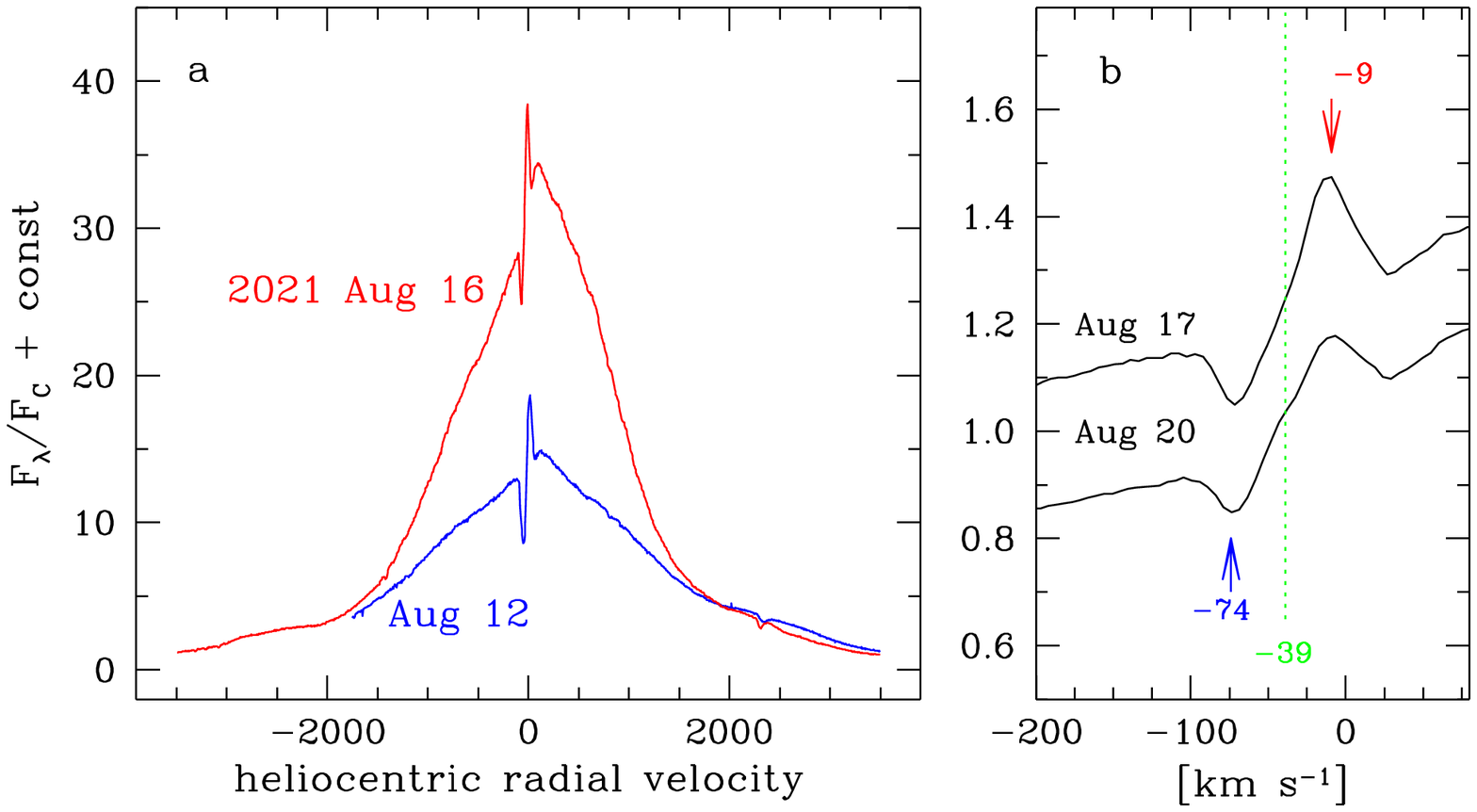}
 \caption[]{ {\bf (a)} H$\alpha$ profiles observed on 12 August 2021 
 and 16 August 2021. A sharp P~Cyg (emission $+$ absorption) component 
 is visible at the top of the strong  emission.  \\
 {\bf (b)}
  Expanded view of the P~Cyg profile. The average positions of the emission ($-9$~km~s$^{-1}$),
  and the absorption ($-74$~km~s$^{-1}$) are marked with arrows. 
  The systemic velocity  ($-39$~km~s$^{-1}$) is marked with vertical green line.
  }
 \label{f2}
\end{figure}
%--------------------------------------------------------------------------------

\subsection{P Cyg profile }

 On Fig.~2a are plotted the H$\alpha$ profiles observed on 12 August 2021 
 and 16 August 2021. 
 A sharp P~Cyg component 
 is visible at the top of the strong emission. 
 It is superimposed on the broad emission line.
 It probably is due to  the outer parts of the slow wind of the red giant 
 and/or the material from the previous outbursts
 ionized by the nova outburst.
 It consists of an blue absorption and red  emission. 
 It is strong and  very well visible on the low resolution spectrum obtained on 12 August 2021.
 We measure the heliocentric  radial velocity of these 
 P~Cyg absorption and emission. They are given in Table~1. 
 The average position 
 of the  P~Cyg absorption is $RV_{abs} = -74 \pm 2$~km~s$^{-1}$
 and 
 of the P~Cyg emission is $RV_{em} = -9 \pm 3$~km~s$^{-1}$.
 The position of the absorption is close to the radial velocity 
 of the central dip of $H\alpha$ at quiescence. 

The analysis of  the CaII and NaD lines of RS~Oph (Patat et al. 2011) 
reveals at least three distinct 
circumstellar components at 
$-77$~km~s$^{-1}$, 
$-63$~km~s$^{-1}$,
and $-46$~km~s$^{-1}$, 
respectively. 
Our measurements indicate that probably,
the component at $-77$~km~s$^{-1}$ is visible in our $H\alpha$ spectra.

 On Fig.~2b is plotted an expanded view of the P~Cyg profile. 
 The average positions of the emission ($-9$~km~s$^{-1}$),
 and the absorption ($-74$~km~s$^{-1}$) are marked with red and blue arrow, respectively. 
 The systemic velocity of RS~Oph is estimated 
 $\gamma = -38.7\pm 0.4$~km~s$^{-1}$  (Brandi et al. 2009)
 and  $\gamma = -40.22 \pm 0.64$~km~s$^{-1}$  (Fekel et al. 2000). 
 The systemic velocity is marked with vertical green line.

 Using the velocities given in Table~1, we can estimated the outflowing velocity, 
 $V_{out}$, as $V_{out} = RV_{em} - RV_{abs} = 64 \pm 4$~km~s$^{-1}$ 
 and $V_{out} = \gamma - RV_{abs} = 34 \pm 2$~km~s$^{-1}$. 
 These two values should be considered as limits, and consecutively 
 our estimate of the outflowing velocity of the material surrounding 
 RS~Oph is 32 km s$^{-1}$ $\le V_{out} \le$ 68~km s$^{-1}$
 (1$\sigma$ error is taken into account). 

Such velocities have been observed from the circumstellar envelopes 
of a few supernovae -- SN~1991T, SN~1998es, SN~2006X (see Fig.~1 of Patat 2013). 
This similarity is a clue that the progenitor systems of some Type Ia supernovae 
can be former recurrent nova systems like RS~Oph (e.g. Patat et al. 2011). 

% Munari \& Valisa (2021, ATel 14840), pointed out that 
% the narrow component is located 60~km~s$^{-1}$ 
% to the blue of the narrow emission component, 
% and could originate in the neutral, external part of the red giant wind.
%
% The very narrow emission component superimposed to the broad emission lines, 
% probably originating in the slow wind of the red giant 
% ionized by the nova outburst. 

\vskip 0.2cm 

{\bf Conclusions:}
We report spectroscopic observations of the recurrent nova RS~Oph obtained 
before and during the 2021 nova outburst. 
For the material surrounding RS~Oph, 
we find outflowing velocity 32~km~s$^{-1}$ $\le V_{out} \le $ 68~km s$^{-1}$,
which is similar to the circumstellar envelopes of some supernovae.
We note that the new GAIA distance indicates that the red giant should be probably
classified in between II and III luminosity class. 

\vskip 0.2cm 

{\bf Acknowledgments:} 
This work is supported by Bulgarian National Science Fund -
projects KP-06-H28/2 "Binary stars with compact object" 
and 
DN~18/13~12.12.2017 "Evolutionary processes in astrophysics: synergy of observations with theory". 
DM  also acknowledges support from the Research Fund of the Shumen University.

\clearpage 
\section{Appendix: Telluric lines around $H\alpha$}
The following telluric lines: 6532.359, 6543.907, 6548.622, 6552.629, 6557.171, 6572.072, 6574.847,
6580.786, 6586.596, 6599.324 are marked on Fig.~3. They are used for tuning the wavelength 
calibration. 
 \begin{figure}
  \vspace{12.7cm}
  \includegraphics{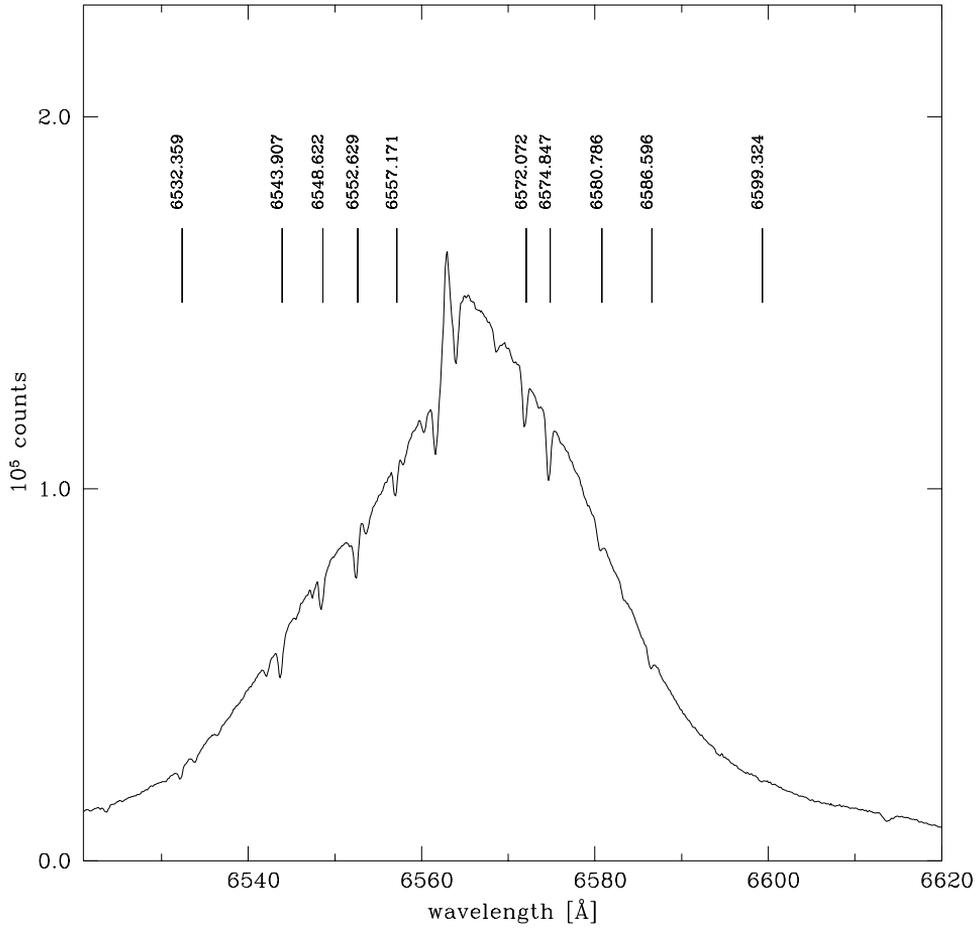}
  \caption[]{Telluric lines imprinted onto the $H\alpha$ emission of RS~Oph. }
   \label{f3}
\end{figure}


\begin{thebibliography}{}
\bibitem{} Bode, M. F. 1987, RS Ophiuchi (1985) and the Recurrent Nova Phenomenon, 
             VNU Science Press, p. 241
\bibitem{} Bode, M.~F., \& Evans, A.\ 2008, Classical Novae, 
             2nd Edition.~Cambridge Astrophysics Series, 
             No.~43, Cambridge: Cambridge University Press, 2008., 43
\bibitem{} Brandi, E., Quiroga, C., Miko{\l}ajewska, J., et al.\ 2009, \aap, 497, 815
\bibitem{} Cheung, C.~C., Ciprini, S., \& Johnson, T.~J.\ 2021, The Astronomer's Telegram, 14834
\bibitem{} Fekel, F.~C., Joyce, R.~R., Hinkle, K.~H., et al.\ 2000, \aj, 119, 1375
\bibitem{} Gaia Collaboration,  Klioner, S.~A.,  Mignard, F.,  et al.\   2021, A\&A, 649, A9
\bibitem{} Gehrz, R.~D., Truran, J.~W., Williams, R.~E., \& Starrfield, S.\ 1998, \pasp, 110, 3
% \bibitem{} I{\l}kiewicz, K., Miko{\l}ajewska, J., Miszalski, B., et al.\ 2019, \aap, 624, A133
\bibitem{} Mikolajewska, J., Aydi, E., Buckley, D., et al.\ 2021, The Astronomer's Telegram, 14852
\bibitem{} Mukai, K.\ 2015, Acta Polytechnica CTU Proceedings, 2, 246 
\bibitem{} Munari, U. \& Valisa, P.\ 2021a, The Astronomer's Telegram, 14840
\bibitem{} Munari, U. \& Valisa, P.\ 2021b,  2021arXiv210901101 
\bibitem{} Patat, F., Chugai, N.~N., Podsiadlowski, Ph., Mason, E., Melo, C., Pasquini, L., 
             2011, A\&A, 530, A63
\bibitem{} Patat, F.,  2013, IAU Symp. 281, "Binary Paths to Type Ia Supernovae Explosions",
             eds. R. Di Stefano, M. Orio, M. Moe, p. 291
\bibitem{} Shafter, A.~W., Rau, A., Quimby, R.~M., et al.\ 2009, \apj, 690, 1148 
\bibitem{} Shidatsu, M., Negoro, H., Mihara, T., et al.\ 2021, The Astronomer's Telegram, 14846
\bibitem{} Straizys, V. \& Kuriliene, G., 1981, Ap\& SS, 80, 353
\bibitem{} Taguchi, K., Ueta, T., \& Isogai, K.\ 2021, The Astronomer's Telegram, 14838
\bibitem{} Tody, D.\ 1993, Astronomical Data Analysis Software and Systems II, 52, 173
\bibitem{} Wagner, S.~J. \& H.~E.~S.~S. Collaboration\ 2021, The Astronomer's Telegram, 14844
\bibitem{} Williams, D., O'Brien, T., Woudt, P., et al.\ 2021, The Astronomer's Telegram, 14849
\bibitem{} Zamanov, R.~K., Boeva, S., Latev, G.~Y., et al.\ 2018, MNRAS, 480, 1363    
\end{thebibliography}
\end{document}